\documentclass[12pt]{article}
\usepackage{epsfig}

\def\beq{\begin{equation}}
\def\eeq#1{\label{#1}\end{equation}}
\def\eeqn{\end{equation}}

\def\beqa{\begin{eqnarray}}
\def\eeqa#1{\label{#1}\end{eqnarray}}
\def\eeqan{\end{eqnarray}}

\let\bar=\overbar

\def\Dslash{\not{\hbox{\kern-4pt $D$}}}
\def\dslash{\not{\hbox{\kern-2pt $\del$}}}

\def\msb{{\bar{\ssstyle M \kern -1pt S}}}

\newcommand{\Om}{$\Omega^{-}$}
\newcommand{\Mo}{$\overline{\Omega}^{+}$}
\newcommand{\Ix}{$\overline{\Xi}^{+}$}
\newcommand{\pt}{$p_{\mathrm{T}}$}
\newcommand{\meanpt}{$\langle p_{\mathrm{T}}\rangle$}
\newcommand{\RAA}{R$_\mathrm{AA}$}

\def\Title#1{\begin{center} {\Large {\bf #1} } \end{center}}

\begin{document}

\Title{Strangeness with ALICE: from $pp$ to Pb-Pb}

\bigskip\bigskip


\begin{raggedright}  

{\it Betty Abelev\index{Abelev, B.} for the ALICE Collaboration\\
Lawrence Livermore National Laboratory\\
Livermore, CA 94550}
\bigskip\bigskip
\end{raggedright}

\section{Introduction}

The physics of relativistic heavy ion collisions is studied to answer questions about primordial quark-gluon matter (QGP) that existed in nature up to $10^{-6}$ sec after the Big Bang.   Today, we can create such matter in high energy heavy ion accelerator-colliders.  The newest, the Large Hadron Collider (LHC), is currently capable of accelerating lead ions to collide at $\sqrt{s_{NN}}=2.76$ TeV, and protons to energy as high as $\sqrt{s}=$ 8 TeV.   One way to characterize the quark chemistry and density, and to investigate particle-production mechanisms in the QGP, is to study the production of the strange quark ($s$) in Pb-Pb and pp collisions.  This is done by identifying and measuring the properties of particles containing one or more $s$-quarks. Comparing the spectra of such particles produced in pp collisions to calculations and phenomenological models can help constrain fragmentation functions; pp spectra also provide a necessary baseline for measurements in Pb-Pb.  In this article, we review the measurements of strange particles performed by the ALICE Collaboration and relate these measurements to matter produced at the LHC. 

\section{Data and the experimental setup}

All data presented here were obtained using the ALICE detector at the LHC.  Details of detector configuration and capabilities of ALICE are described elsewhere \cite{ALICE}.  In Pb-Pb, collision centrality was determined using the signals from VZERO detector and tracklets obtained in the Inner Silicon Tracking (ITS) system.  Strange and multistrange particle spectra were measured using tracks reconstructed in the ITS and the Time Projection Chamber (TPC).  Proton, pion, and kaon tracks were identified using particle energy loss in the TPC.  Using topological considerations, $\Lambda$ ($\Lambda\rightarrow\pi^-+p^+$), $\Xi^-$ ($\Xi^-\rightarrow\Lambda+\pi^-$) and \Om\ ($\Omega^-\rightarrow\Lambda+K^-$)  baryons and their anti-particles, as well as $\mathrm{K^0_S}$ mesons ($\mathrm{K^0_S}\rightarrow\pi+\pi$) were reconstructed via the corresponding decay channels, as described, for example, in \cite{ALICE_900Str}.   After reconstruction, the spectra were corrected for acceptance and detector effects, and normalized to the inelastic event cross-section.

\begin{center}
\begin{figure}[ht]
\begin{minipage}[t]{0.46\linewidth}
\centering
\epsfig{file=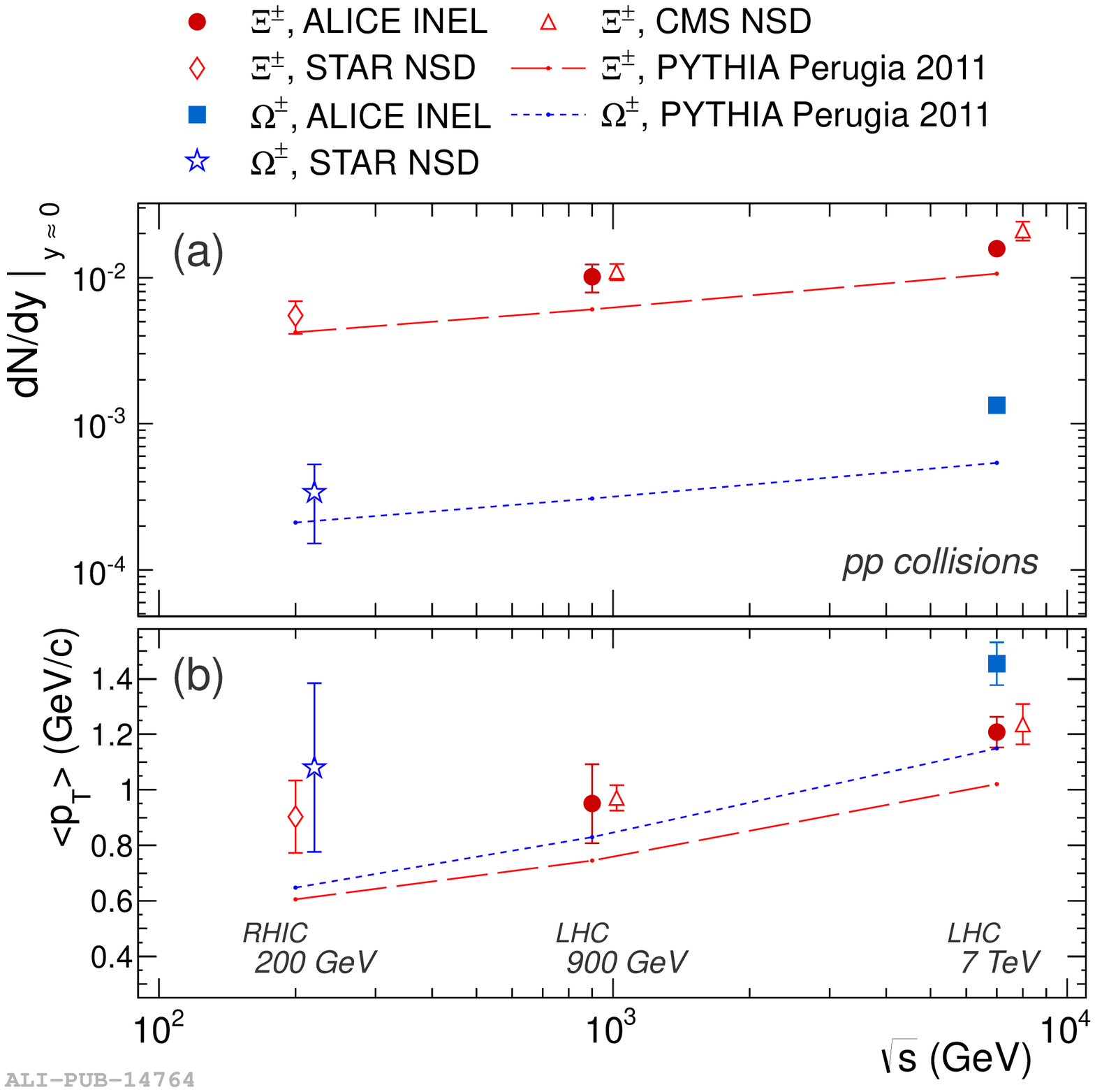,height=2.3in}

\caption{$\Omega^{\pm}$ and $\Xi^{\pm}$ yields (a) and \meanpt\ (b) as a function of collision energy in pp collisions.}
\label{fig:ppTrend}
\end{minipage}
\hspace{0.2in}
\begin{minipage}[t]{0.46\linewidth}
\centering
\epsfig{file=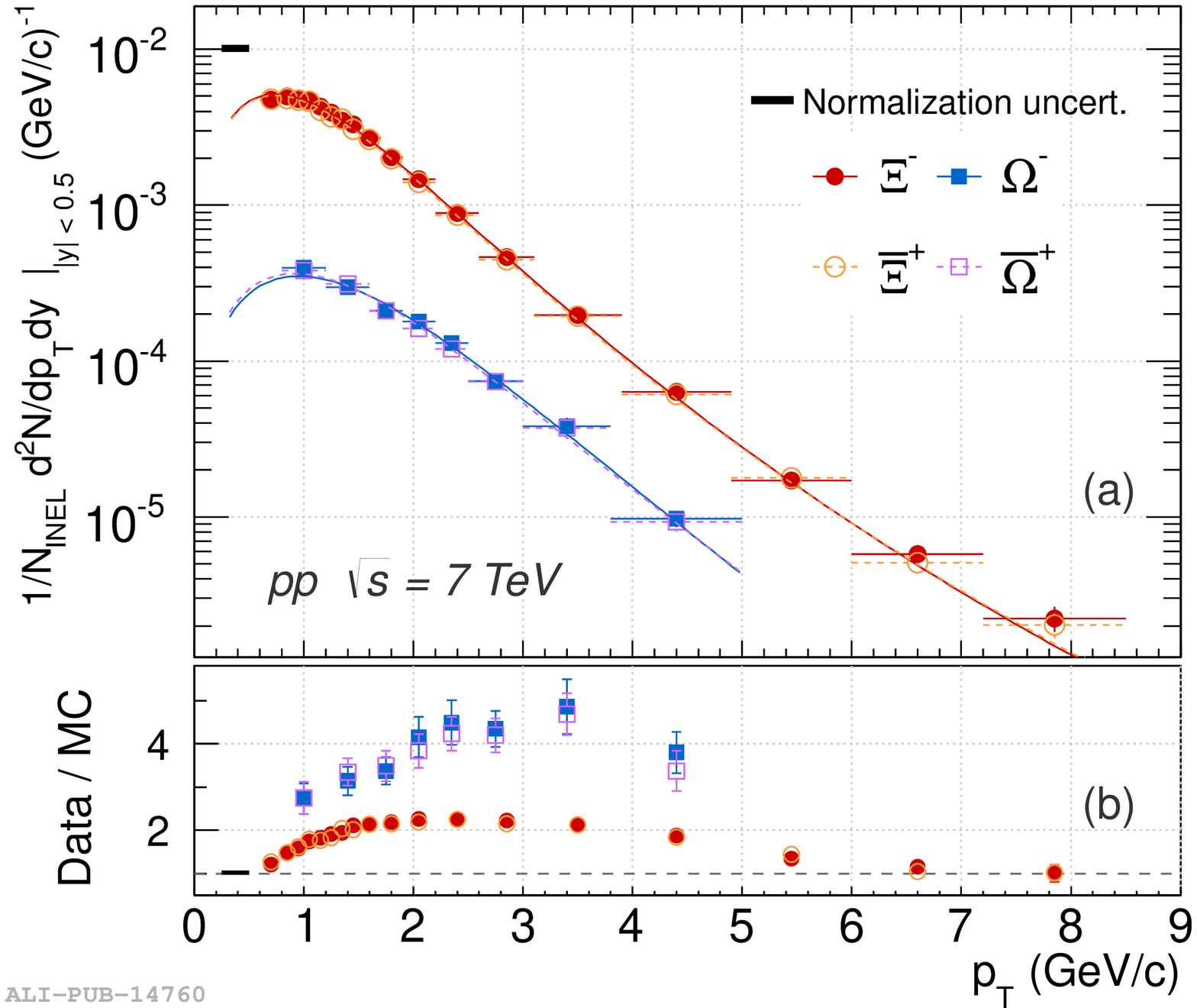,height=2.2in}
\caption{7 TeV pp multi-strange spectra (a) and PYTHIA Perugia-2011 comparison (b).}
\label{fig:ppSpec}
\end{minipage}
\end{figure}
\end{center}
\begin{center}
\begin{figure}[ht]
\begin{minipage}[t]{0.46\linewidth}
\centering
\epsfig{file=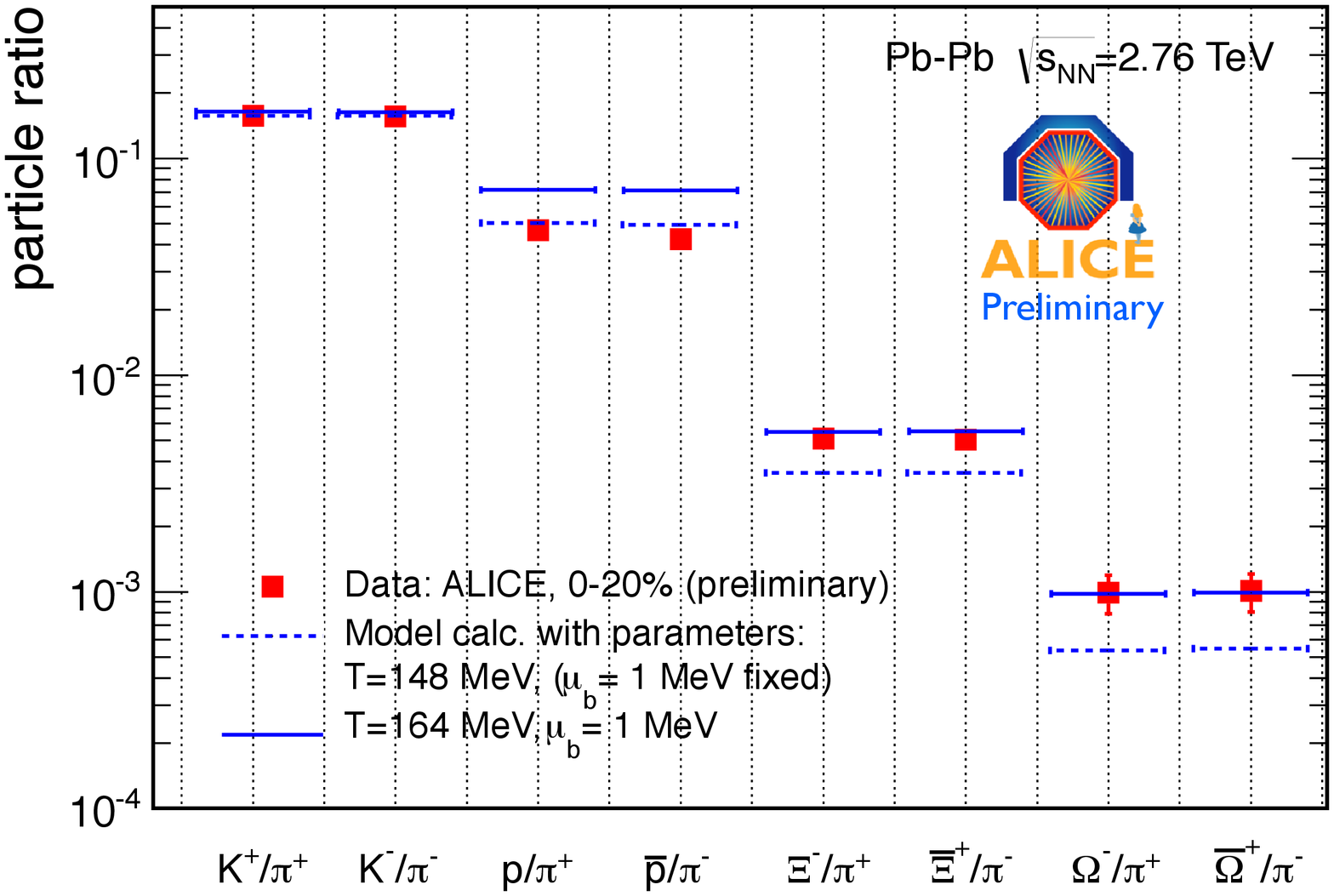,height=1.9 in}
\caption{Particle ratios measured by ALICE in 2.76 TeV Pb-Pb collisions, and two thermal model fits, at T=148 MeV (dashed line), and at T=164 MeV (solid line).  $\gamma_s$ and $\mu_B$ are set to 1.}
\label{fig:TherMod}
\end{minipage}
\hspace{0.2in}
\begin{minipage}[t]{0.46\linewidth}
\centering
\epsfig{file=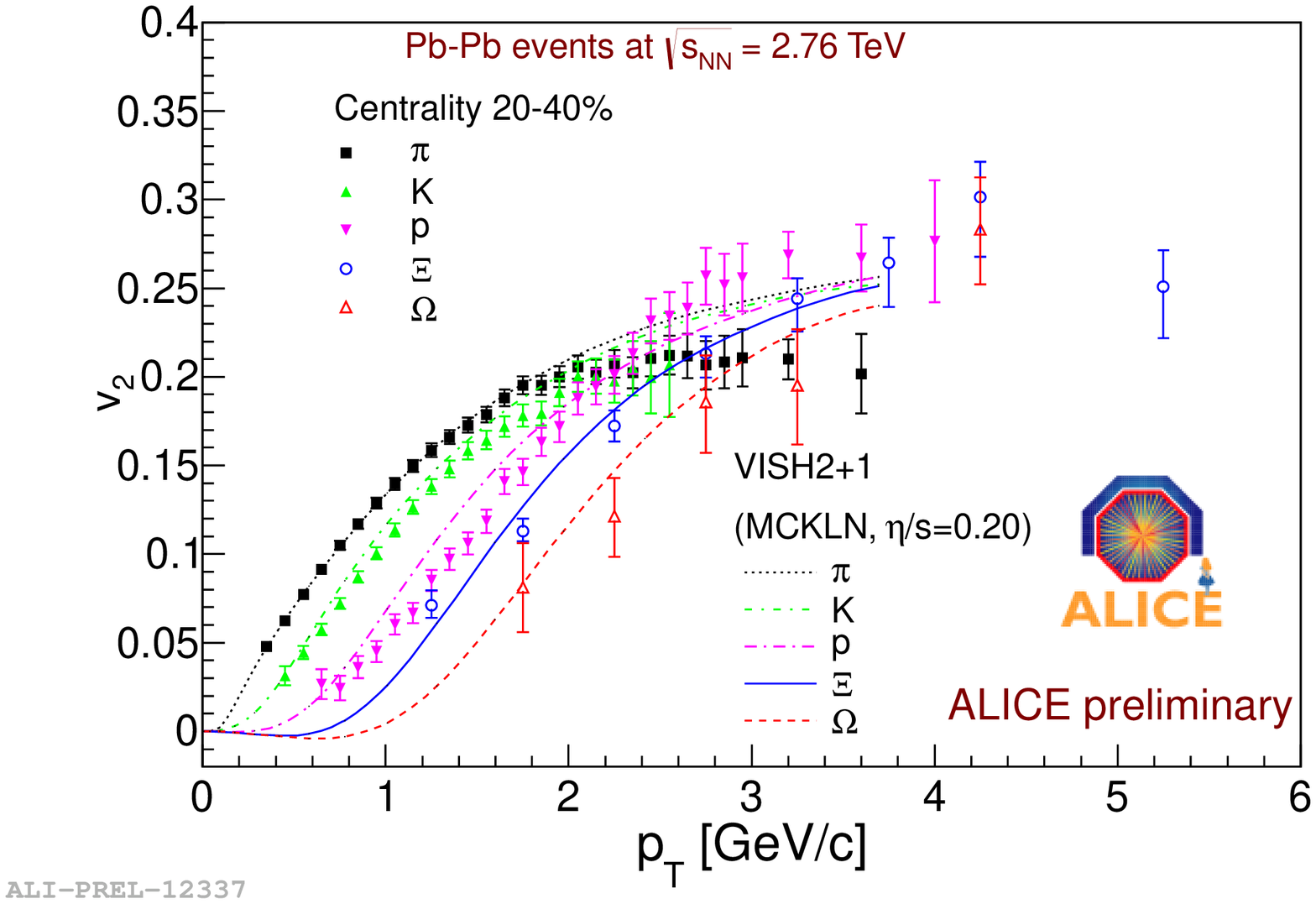,height=1.9in}

\caption{$v_2$ for pions, protons, kaons, and $\Xi$ and $\Omega$ baryons in 20-40\% central Pb-Pb collisions (symbols).  Also plotted are VISH2+1 with $\eta/$s=0.2 curves for the same particles.}
\label{fig:v2}
\end{minipage}
\end{figure}
\end{center}
\vspace{-1in}
\section{Results}
ALICE measured multi-strange baryon spectra in pp at two energies, $\sqrt{s}=$ 0.9 TeV and 7 TeV.  The trends, together with a 0.2 TeV measurement by STAR \cite{STARpp}, are shown in Fig.~\ref{fig:ppTrend}.  To increase statistics, the particle and the anti-particle yields and \meanpt\ are added together, and denoted as  $\Omega^{\pm}$ (for \Om\ and \Mo) and $\Xi^{\pm}$ (for $\Xi^-$ and \Ix).
The $\Omega^{\pm}$ and  $\Xi^{\pm}$  spectra at 7 TeV were also compared to several PYTHIA tunes \cite{multiStr}.  The  PYTHIA Perugia-2011 tune \cite{PYTHIAtunes} was the best match to our spectra.  The comparison between model and experimental data is shown in panel (b) of Fig.~\ref{fig:ppSpec}.  Perugia-2011 differs from other PYTHIA tunes in that the pop-corn meson production mechanism is turned off.

\begin{center}
\begin{figure}[ht]
\begin{minipage}[t]{0.46\linewidth}
\centering
\epsfig{file=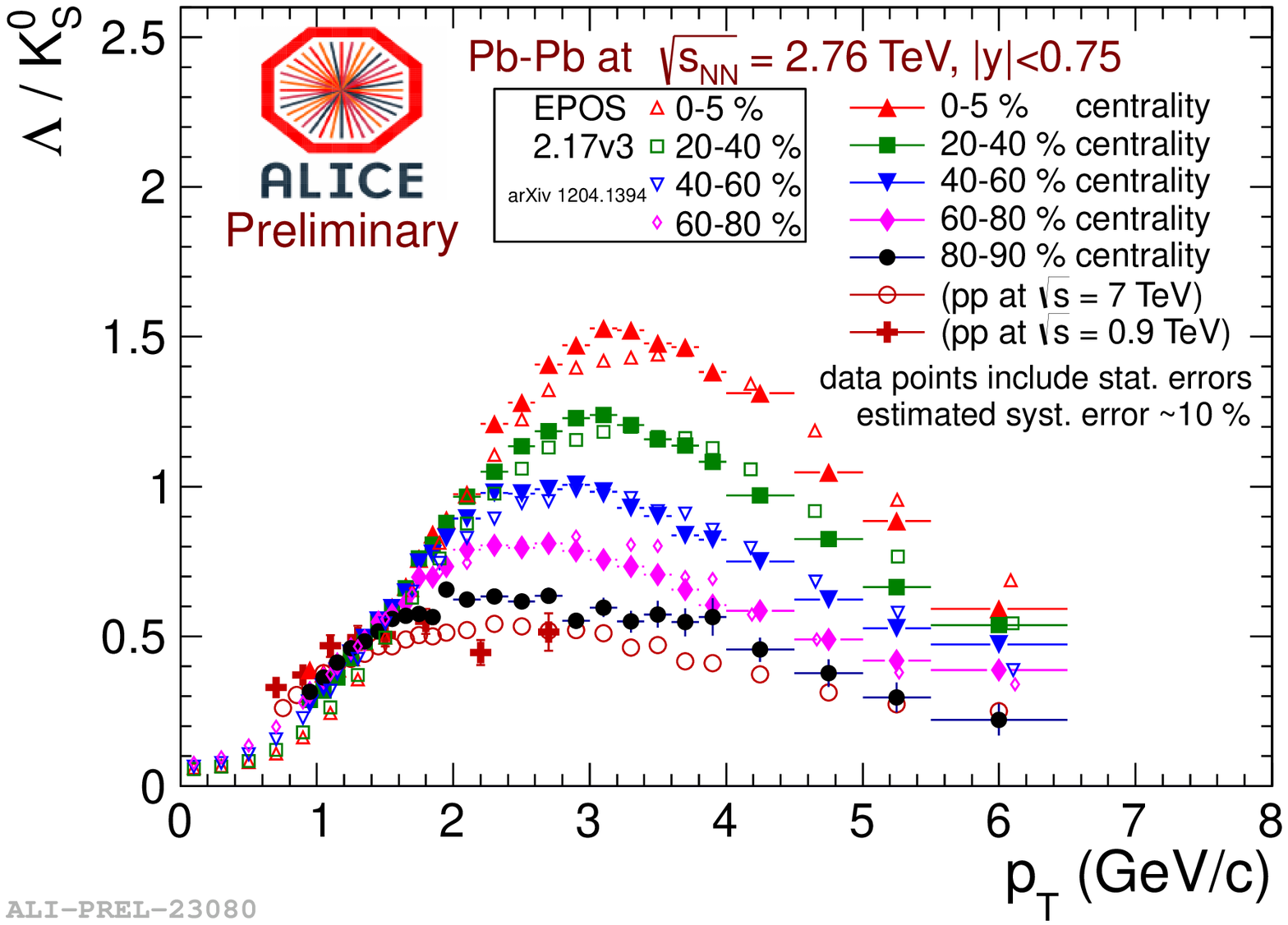,height=1.9 in}
\caption{$\Lambda/\mathrm{K}^0_\mathrm{S}$ as a function of \pt\ at all 2.76 TeV Pb-Pb collision centralities, and in 0.9 and 7 TeV pp collisions.}
\label{fig:lamOverK0}
\end{minipage}
\hspace{0.2in}
\begin{minipage}[t]{0.46\linewidth}
\centering
\epsfig{file=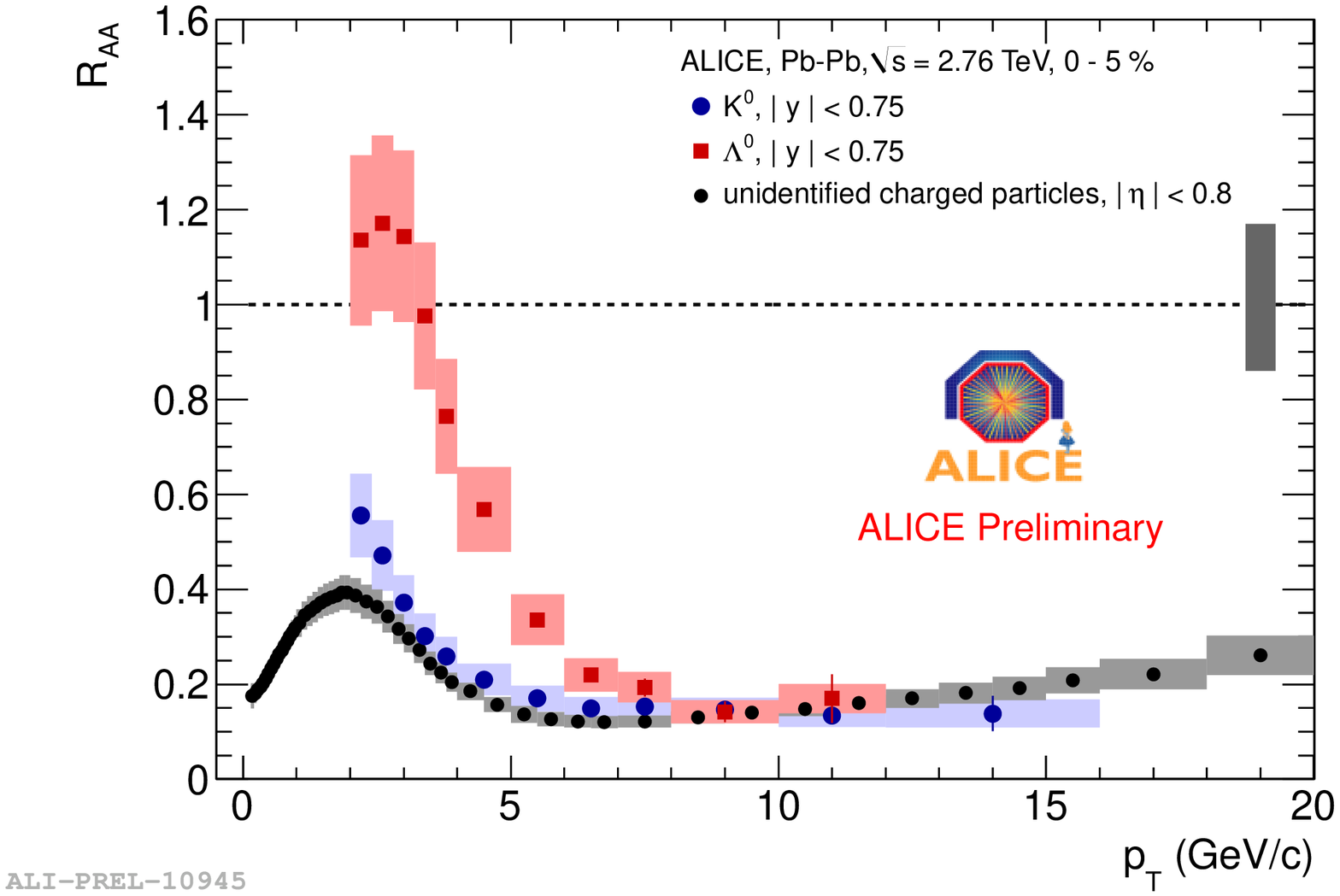,height=1.9in}

\caption{\RAA, for K$^0_\mathrm{S}$, $\Lambda$ and charged particle spectra in 2.76 TeV Pb-Pb 0-5\% central collisions.}
\label{fig:Raa}
\end{minipage}
\end{figure}
\end{center}

\begin{figure}[htb]
\begin{center}
\epsfig{file=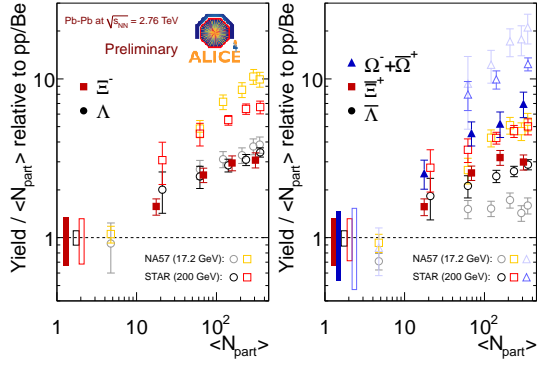,height=1.9in}
\caption{Strange particle production as a function of $\langle\mathrm{N}_\mathrm{part}\rangle$ for $\sqrt{s_{NN}}=$ 0.017, 0.2, and 2.76 TeV collisions relative to pBe (NA57) and pp (STAR, ALICE).}
\label{fig:Enhancement}
\end{center}
\end{figure}

In Pb-Pb collisions, a broad range of measurements were made.  Strange particle yields, together with non-strange pions and protons, were fit to a thermal model \cite{thermalMod}, as seen in Fig.~\ref{fig:TherMod}.   The elliptic flow coefficient, $v_2$, for strange and multi-strange particles was determined, and is shown in Fig.~\ref{fig:v2} together with low-viscosity ($\eta/$s=0.2) VISH2+1 hydrodynamical calculations \cite{hydro}.  To get a handle on the \pt\ range at which particle production via coalescence is applicable, $\Lambda/\mathrm{K}^0_\mathrm{S}$ ratios were constructed for all Pb-Pb centralities and for the two pp data sets, as seen in Fig.~\ref{fig:lamOverK0}. In Fig.~\ref{fig:Raa}, the nuclear modification ratio, \RAA, of strange particles is shown together with the \RAA\ of all charged particles measured by ALICE.  Finally, in Fig.~\ref{fig:Enhancement}, we show the enhancement in the production of particles with $s$-quarks with respect to baseline collisions (pBe at NA57, and pp in others)  as a function of collision centrality (characterized by N$_\mathrm{part}$, number of participant nucleons) and the $s$-quark content.  

\section{Summary and conclusions}
Particles containing the $s$-quark are a multi-faceted probe, used in relativistic heavy ion collisions with great success to test low to high \pt\ regimes, help measure fragmentation functions in pp collisions, and characterize the collective properties of the medium in Pb-Pb events. ALICE measurements have validated the PYTHIA Perugia-2011 tune's removal of the pop-corn meson creation mechanism, since it improved significantly the description of the multi-strange data in 7 TeV pp collisions \cite{multiStr}.   In Pb-Pb collisions, the strange quark seems to be thermalized at T=164 MeV, when $\gamma_s$ is set to 1 (i.e., strangeness production is saturated).  We also measure a large volume enhancement in the multi-strange particle production, consistent with previous observations and predictions \cite{StrangeEnh}.   Particles with more $s$-quarks experience a greater enhancement with respect to the baseline.  However, another trend is confirmed -- the amount of enhancement decreases with increased collision energy, most likely due to the power-law increase in baseline yields as collision energy increases.    Collective effects in Pb-Pb collisions are consistent with light-flavour non-strange particle observations and also with those observed at lower energies.  At 2.76 TeV, the strange-particle $v_2$ measurements are consistent with a low-viscosity medium, the meson-baryon ratios at intermediate \pt\ point to the dominance of recombination in that region, and at high \pt\ the strange particles seem to be suppressed as much as charged particles.

\def\Discussion{
\setlength{\parskip}{0.3cm}\setlength{\parindent}{0.0cm}
     \bigskip\bigskip      {\Large {\bf Discussion}} \bigskip}
\def\speaker#1{{\bf #1:}\ }
\def\endDiscussion{}
%




 
\end{document}